\documentclass[twocolumn,showpacs,superscriptaddress, preprintnumbers , amsmath,amssymb,letterpaper]{revtex4}

\usepackage{graphicx}
\usepackage{dcolumn}
\usepackage{bm}
\usepackage{amsmath}
\usepackage{multirow}

\newcommand{\mean}[1]{\mbox{$\langle{#1}\rangle$}}

\usepackage[colorlinks=true]{hyperref}

\begin{document}

\title{Amp\`ere-Class Pulsed Field Emission from Carbon-Nanotube Cathodes \\
in a Radiofrequency Resonator}

\author{D. Mihalcea}
\affiliation{Northern Illinois Center for Accelerator \& Detector Development and Department of Physics, Northern Illinois University, DeKalb, IL 60115, USA}
\author{L. Faillace}
\affiliation{Radiabeam Technologies LLC, Santa Monica, CA 90404, USA}
\author{J. Hartzell}
\affiliation{Radiabeam Technologies LLC, Santa Monica, CA 90404, USA}
\author{H. Panuganti}
\affiliation{Northern Illinois Center for Accelerator \& Detector Development and Department of Physics, Northern Illinois University, DeKalb, IL 60115, USA}
\author{S. M. Boucher}
\affiliation{Radiabeam Technologies LLC, Santa Monica, CA 90404, USA}
\author{A. Murokh}
\affiliation{Radiabeam Technologies LLC, Santa Monica, CA 90404, USA}
\author{P. Piot}
\affiliation{Northern Illinois Center for Accelerator \& Detector Development and Department of Physics, Northern Illinois University, DeKalb, IL 60115, USA}
\affiliation{Fermi National Accelerator Laboratory, Batavia, IL 60510, USA}
\author{J. C. T. Thangaraj}
\affiliation{Fermi National Accelerator Laboratory, Batavia, IL 60510, USA}

\date{\today}
\begin{abstract}
Pulsed field emission from cold carbon-nanotube cathodes placed in a radiofrequency resonant cavity was observed. The cathodes were located on the backplate of a  conventional $1+\frac{1}{2}$-cell resonant cavity operating at 1.3-GHz and resulted in the production of bunch train with maximum average current close to 0.7 Amp\`ere. The measured Fowler-Nordheim characteristic, transverse emittance, and pulse duration are presented and, when possible, compared to numerical simulations. The implications of our results to high-average-current electron sources are briefly discussed. \end{abstract}

\preprint{FERMILAB-PUB-14-527-APC}
\maketitle

Over the last decades field-emission (FE) $-$ the emission of electrons via tunneling effect $-$ has led to the development of compact electron sources that have been widely disseminated in microelectronics~\cite{HP}, electron microscopy~\cite{SEM}, and more recently as particle-accelerator sources~\cite{brau}.  FE enjoys a greater simplicity 
compared to other types of electron-emission mechanisms: it does not require auxiliary systems such as lasers employed in photoemission nor needs to be electrically heated  
by a filament as for thermionic cathode.  A FE cathode operates with intense [${\cal O}\mbox{(GV/m)}$] electric fields applied at its surface. The field bends the potential barrier of the material and enhance the probability for electron tunneling. In practice, given the limited electric-field amplitude sustainable in common apparatus ($10-100$~MV/m), the generation of large field relies on field enhancement provided by sharp microscopic features present at the surface of the FE cathode. Such attribute results in the local field 
in the vicinity of the features  $E_e=\beta_e E$ where $\beta_e$ is the enhancement factor and $E$ is the applied (macroscopic) field. Consequently the current 
density emitted from one of the sharp features is governed by the Fowler-Nordheim's (FNÕs) law~\cite{FN}  $ {\pmb j}=a E_e^2\exp\left(-\frac{b}{E_e}\right) {\hat{\pmb n}}$, where $a\equiv \frac{1.42\times10^{-6}}{\Phi}\exp\left(\frac{10.4}{\Phi^{1/2}}\right) $ and  $b\equiv -6.56 \times 10^9\Phi^{3/2}$ are constants that depend on the material work function $\Phi$ (in units of eV)~\cite{minoux},  ${\hat{\pmb n}}$ is the unitary vector normal to the local emitting surface. FE cathodes consisting of single field emitter have been proposed as source of ultra-bright electron bunches~\cite{Hommelhoff}. Conversely, FE cathodes composed of a large number of field emitters are contemplated as high-current electron sources that could be deployed in a variety of contexts ranging from fundamental science along with medical, industrial, and security settings. Finally, the application of a time-dependent electric field results in the generation of electron bunches with finite duration~\cite{IEEEpaper,nature,piotAPL,UDCAPL}. 
Field emission from various materials have been  extensively studied and most recently the use of carbon nanotube (CNT) has significantly increased~\cite{minoux}. CNTs are allotropes of carbon with a cylindrical nanostructure and with exceptional electrical and mechanical properties. CNT's are fibers with diameters  ranging from 1 to 50 nm, coming in either single-wall or multi-wall strands and with aspect ratio of up to $\sim 1000$~\cite{zhu}. These properties yield substantial field enhancement factors and alongside with their low electrical resistance, high thermal stability and robustness to high temperatures, CNTs are excellent candidates for FE cathodes. CNTs can be synthesized as aligned field-emitter arrays (FEA) or configured as randomly oriented field emitters deposited on surfaces.  Although FEAs are ideal for most applications, especially in vacuum microelectronics~\cite{fursey}, randomly oriented CNTs  enjoy a simpler fabrication process and can be formed on shaped surfaces. The latter capability could enable the generation of transversely-tailored high-current electron beams such as needed  for, e.g., electron-beam-based manipulations of high-intensity ion-beams accelerators~\cite{iota}. \\
\begin{figure}[hhhhhh!!!!!!!!!!!!]
\centering
\includegraphics[width=0.45\textwidth]{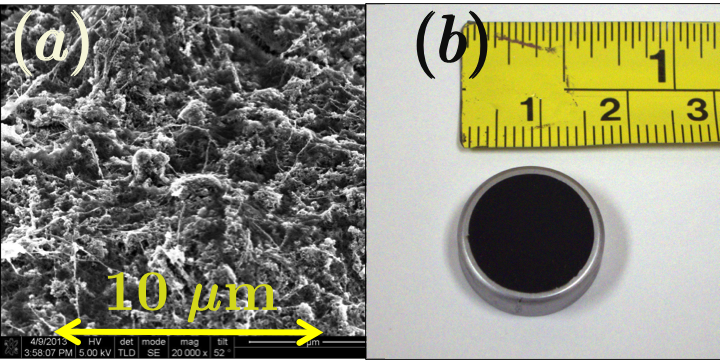}
\caption{Micrograph of CNT cathode surface (a) and photograph of the CNT cathode on its molybdenum substrate (b). \label{fig:cntpic}}
\end{figure}

In this Letter, we report on the experimental tests of pulsed emission from CNT cathodes.  The cathodes were synthesized using an electrophoretic deposition (EPD) process and consisted of a layer of multiple allotropes of carbon including  nanotubes, buckyballs, graphite, and amorphous carbon~\cite{hartzell}; see Fig.~\ref{fig:cntpic}(a).  Emerging from this layer are a vast number of randomly-oriented nanotubes. Both the extremities and sides of the nanotubes can act as field emitters. The current formed by such cathodes can be obtained as the surface integration of the current density over the CNT deposition area and is written as $I= {\cal A} j$ for a flat surface where ${\cal A}$ is the effective emission area. Two cathodes were tested, a ``large"  and a "small" cathodes which respectively consisted of a 15-mm and 1.5-mm diameter CNT-coated circular area deposited on respectively a molybdenum and stainless steel substrate; see Fig.~\ref{fig:cntpic}(b).

\begin{figure}[hhhhh!!!!!!!!!!!!]
\centering
\includegraphics[width=0.45\textwidth]{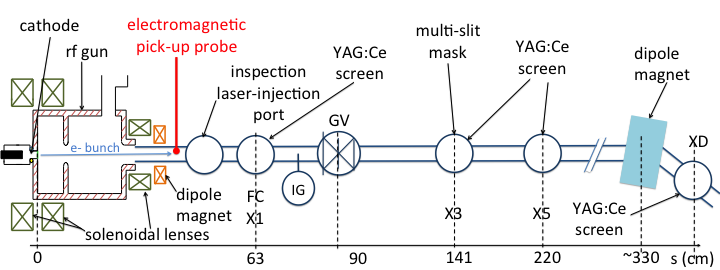}
\caption{Top-view schematics of the experimental setup of the HBESL facility. The "X's" label indicate the location of diagnostics, ``FC", ``IG" and ``GV" respectively stands for 
``faraday cup", ``ion gauge", and vacuum ``gate valve". Only beam-line elements pertinent to the experiment are shown. }\label{fig:expsetup}
\end{figure}

The experimental characterization of the cathodes was carried at the high-brightness electron beam source (HBESL) located at Fermilab~\cite{carneiro}. The facility incorporates a radiofrequency (RF) gun followed by a beam line instrumented with various beam diagnostics depicted in Fig.~\ref{fig:expsetup}. The RF gun is a 1+1/2 cell resonant cavity operating on the TM$_{010,\pi}$ mode at $f_0=1.3$~GHz and is powered by a pulsed klystron capable of producing up to 2 MW of peak power. For the experiment reported in this Letter, the klystron was operated at 1 Hz with a pulse duration of 30~$\mu$s. The gun is nested in three magnetic lenses (referred to as ÒsolenoidsÓ) that are nominally used to control the beam divergence and transverse emittance. The two cathodes described above were mounted on a standard cathode-plug holder and inserted in the RF gun.   
The available diagnostics along the downstream accelerator beam line includes a Faraday cup, several transverse-density monitors consisting of remotely insertable Cerium-doped Yttrium Aluminum Garnet (YAG:Ce) scintillators, and a set of capacitive electromagnetic pick-ups that can detect the transient electromagnetic field produced by the passing electron bunches.  Several dipole magnets are used to deflect the beam and infer its mean momentum. Finally, the beam transverse emittance $\varepsilon_u \equiv \frac{1}{m_e c}[\mean{u^2}\mean{p_u^2}-\mean{up_u}^2]^{1/2}$ can be measured using a multi-slit method~\cite{lejeune} $-$ here   $(u,p_u)$ refers to the position-momentum coordinate along the horizontal ($u=x$) or vertical ($u=y$) degree of freedom,  $\mean {.}$ represents the statistical averaging over the beam phase space distribution, and $m_e$ and $c$ are the electron rest mass and the velocity of light.  



\begin{figure}[hhhhh!!!!!!!!!!!!]
\centering
\includegraphics[width=0.5\textwidth]{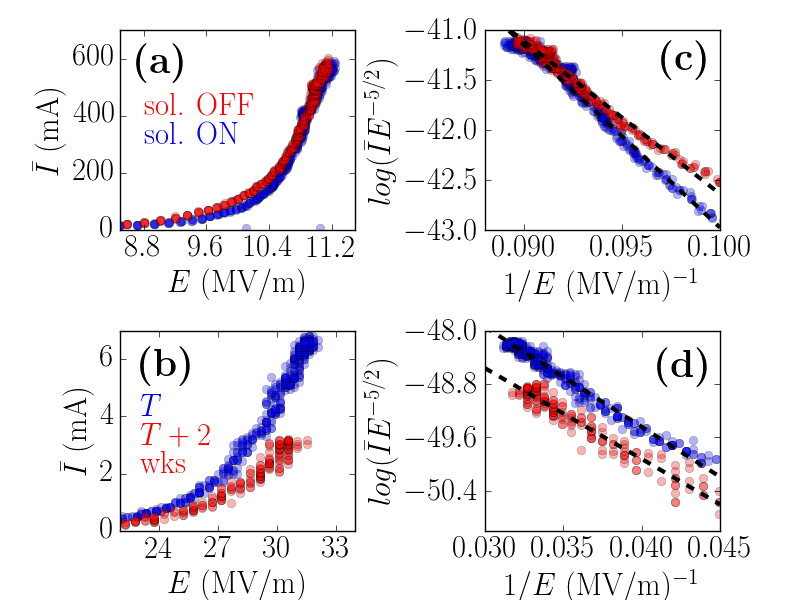}
\caption{Measured average current $\bar{I}$ as a function of applied microscopy field (a,b) and corresponding Fowler-Nordheim plots (c,d).  The upper and lower rows respectively correspond to the large and small cathodes. The dashed lines in (c,d) represent linear polynomial fits. For plots (b,d) the blue and red symbols respectively correspond to data taken just after installation of the small cathode and 2 weeks later.}\label{fig:FNplots}
\end{figure}

To first characterize the field-emission process, the emitted current versus applied macroscopic field was measured for different conditions. The current 
inferred from the Faraday cup is time averaged and its functional dependence on the applied field is given by 
$\bar{I}= \frac{1}{\sqrt{2\pi}}{\cal A} a (\beta_e E)^{5/2} \exp\left(-\frac{b}{\beta_e E}\right)$~\cite{FNRF}. The  $\bar{I}-E$ curves display the 
expected exponential dependence of the current; see Fig.~\ref{fig:FNplots}[(a), and (c)]. Furthermore when reported on a FN diagram $[1/E,\log(\bar{I} E^{-5/2})]$, 
the data appear as lines and a linear-regression analysis provides information on the averaged enhancement factor and effective emission area as summarized in 
Table~\ref{tab:beta} for four of the cases studied. 
The settings of the three lenses were varied simultaneously and set to insure a zero magnetic field on the cathode surface and gave rise to a $\sim 10$\%  relative variation in produced beam current confirming a significant fraction of  the current is actually transversely captured and transported up to the location of the Faraday cup. We found the values of the enhancement factors to be independent of the applied magnetic field [Fig.~\ref{fig:FNplots}(a,b)] and to be qualitatively similar for the two cathodes used during our experiments; see Fig.~\ref{fig:FNplots}(c,d).  It should be noted that the effective area ${\cal A}$ is much smaller than expected for the small-area cathode. 
\begin{table}[hhh!]
\caption{\label{tab:beta} Inferred enhancement factor $\beta_e$ and effective emission area ${\cal A}$ for the four operating cases considered in
the text. The values are written as $A^{+u}_{-v}\pm w $, where $A$ is obtained for a nominal work function value $\Phi=4.9$~eV while 
the upper and lower uncertainties $u$ and $v$ are respectively evaluated for $\Phi=5.4$ and $4.5$~eV.  The error bar $w$ is propagated from the uncertainty 
on the linear regression. }
\begin{center}
\begin{tabular}{l c c c }
\hline \hline  configuration         &   &   $\beta_e$       & ${\cal A}\times 10^{16}$ (m$^2$)  \\ \hline
small cath., B field off     &    &  $468.1^{+73.4}_{-69.8} \pm 4.7$      & $5.15^{+0.15}_{-0.22} \pm 0.10$   \\
small cath., B field off\footnote{data taken 2 weeks later than data on previous line.}     &    &  $ 504.6^{+79.2}_{-75.2} \pm 5.1$     & $ 1.57^{+0.05}_{-0.07} \pm 0.03 $   \\
\hline                &   &   $\beta_e$       & ${\cal A}\times 10^{7}$ (m$^2$)  \\ \hline
large cath., B field off      &   &  $ 395.3^{+62.0}_{-58.9} \pm 4.0$      & $ 10.18^{+0.29}_{-0.44} \pm 37.62 $   \\
large cath., B field on      &  &  $ 468.1^{+73.4}_{-69.8} \pm 4.7$       & $ 0.56^{+0.02}_{-0.02} \pm 2.81 $   \\
\hline \hline
\end{tabular}
\end{center}
\end{table}
Assuming the same CNT density for both cathode we would anticipate the effective emission area associated to the small cathode to be  $(1.5/15)^2=10^{-2}$ while a factor $\sim [10^{-10}-10^{-8}]$ is observed. A post-experiment inspection of the cathodes indicated some damages (dark spots) on the small cathode attributed to multipacting occurring due to the favorable secondary-emission yield of the stainless-steel substrate (strong multipacting emission was observed during operation of the small cathode). The small cathode also consistently degraded with time when operated at high field. The large cathode did not show any performance degradation despite being exposed to atmosphere for  $\sim 4$ weeks between two subsequent tests. \\
\begin{figure}[tttt!!!!!!!!!!!!]
\centering
\includegraphics[width=0.475\textwidth]{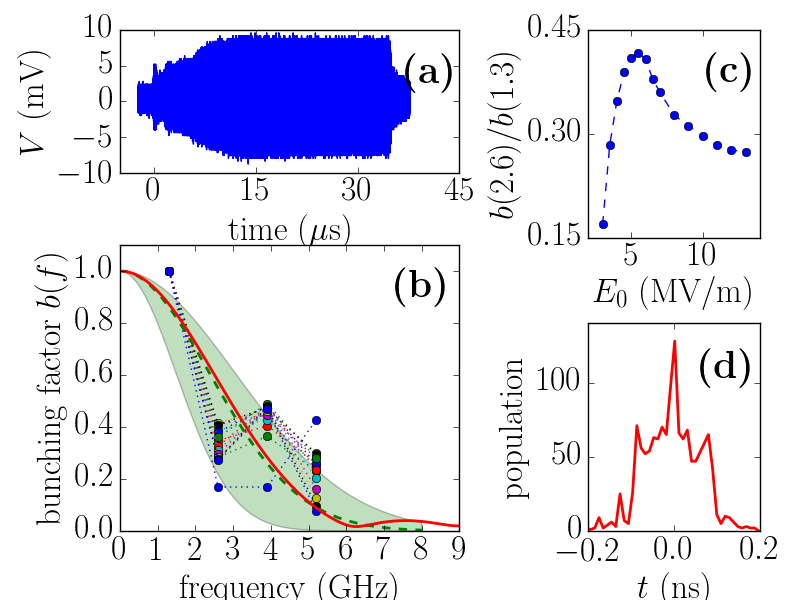}
\caption{Voltage detected from the electromagnetic pickup (a), corresponding bunching factor (b) and evolution of the bunching factor evaluated at $f=2.6$~GHz as function of $E_0$ (c). In (b) the data points are FFTs of different traces obtained for different values of $E_0$, the dash line represents a fit considering to a Gaussian bunch distribution (the shaded green area accounts for the uncertainties in the fit) and the solid thick line corresponds to the simulated bunch distribution using the {\sc warp} program shown in (d). }\label{fig:BPM}
\end{figure}
%
The beam temporal structure of the electron beam formed by the large cathode was characterized using an electromagnetic pick-up located 30 cm from the cathode. The transient voltage induced by the bunches was detected by a capacitive coupler and recorded on a 12-Gs oscilloscope; a typical trace is displayed in Fig.~\ref{fig:BPM}(a). The detected signal can be factored as $V(t) = e(t)\sum_{n=1}^N \Lambda (t + nf_0^{-1})$  where $e(t)$ is the signal envelope, $\Lambda(t) \propto i(t)$ is the signal induced by one bunch. The amplitude of the fast-Fourier transform (FFT) of $V(t)$   provides the bunching factor $b(f)$ which is enhanced at harmonics of the bunch repetition frequency $f_0$ as shown in Fig.~\ref{fig:BPM}(b). The relative amplitudes  $b(f)/b(f_0)$ at  harmonic frequencies $f=nf_0$ (with $n\ge 2$) provide an upper bound for the bunch duration. Analysis of the data presented in Fig.~\ref{fig:BPM}(b), assuming the electron bunch follows a Gaussian temporal distribution,  gives an rms bunch duration of $\sigma_t \simeq 67\pm 25$~ps at an applied field $E \in [5,12]$~MV/m. The electron-bunch  duration is expected to scale as $\sigma_t \propto \sqrt{E}$~\cite{piotAPL} at the cathode. The bunch length increase with $E_0$ is supported by the observed increase of the $b(2f_0)/b(f_0)$ [see Fig.~\ref{fig:BPM}(c)] but due to the limited resolution of our pulse-length-measurement technique and the complicated dynamics in the RF gun the functional dependence $\sigma_t(E)$ could not be characterized. In spite of these limitations, the measured pulse duration agrees reasonably well with particle-in-cell simulations performed with the {\sc warp} framework~\cite{vay} which includes a self-consistent field-emission model~\cite{alex}; see Fig.~\ref{fig:BPM}(b,d). The FE parameters used in the simulations are the one reported for the large cathode (solenoid off) in Table~\ref{tab:beta}. 
\begin{figure}[bb!!!!!!!!!!!!]
\centering
\includegraphics[width=0.495\textwidth]{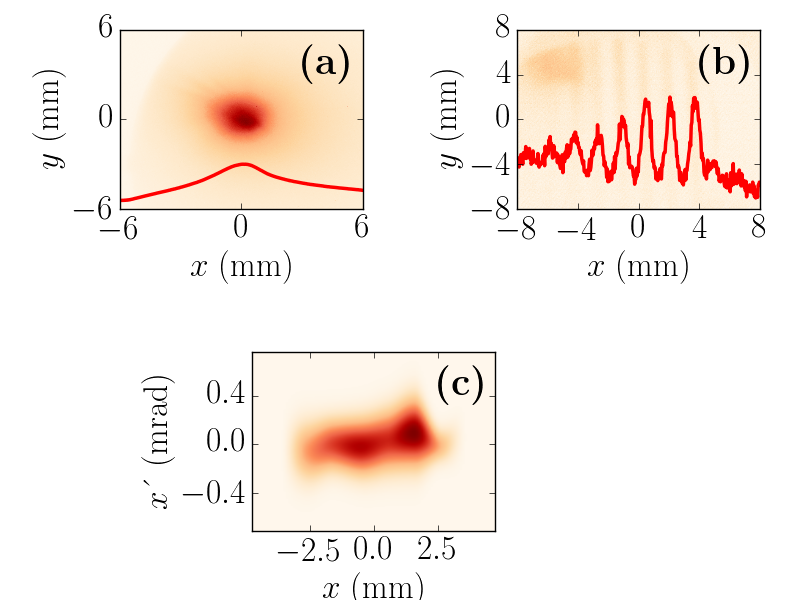}
\caption{Emittance measurement snapshots showing the beam transverse distribution at X3 (a), the transverse distribution of the beamlets transmitted through the multislit mask observed at X5 (b) with associated horizontal projections (red traces). Image (c) shows the reconstructed horizontal  $(x,x'\equiv p_x/pz)$ trace space at the location  of X3 from processing of images (a) and (b). These measurements were performed for the small cathode.}\label{fig:emittance}
\end{figure}

An important figure of merit of the field-emitted beam is its transverse emittance. The horizontal emittance of the full bunch train was characterized for the small cathode. The multislit mask located a position X3 was inserted and  the transmitted beamlets were observed at location X5. A measurement of the beamlet root-mean-square (rms) size at X5 provides information on the beam intrinsic divergence $\sigma'_u$ at X3. Together with a measurement of the rms transverse beam size $\sigma_u$ at X3, the divergence yield the value of the transverse normalized emittance as $\varepsilon_u = \beta\gamma \sigma'_u \sigma_u$ where $u \in[x,y]$ refers to one of the transverse degrees of freedom and $\beta\equiv (1-\gamma^{-2})^{1/2}$ where $\gamma$ is the relativistic Lorentz factor.  An example of measurement with reconstructed phase space appears in Fig.~\ref{fig:emittance}. The measurement indicates a transverse horizontal emittance of  $\varepsilon_x = 2.64 \pm 0.8$~$\mu$m for the small cathode. Measurements for the large cathode were compromised by the large energy spread.

\begin{figure}[tt!!!!!!!!!!!!]
\centering
\includegraphics[width=0.5\textwidth]{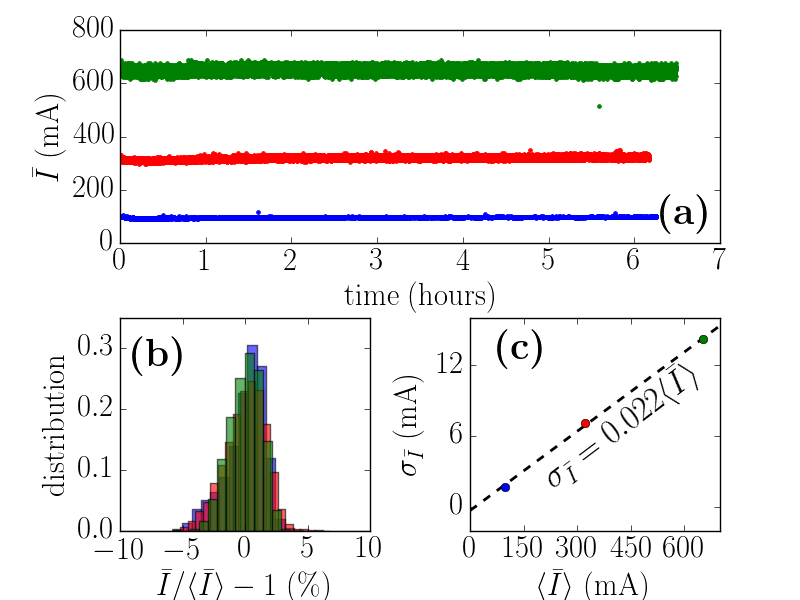}
\caption{Current evolution over a $>6$-hour period (a) for 100 (blue) , 300 (red), and 650 mA (green) with corresponding histograms (b) and rms fluctuations (c). }\label{fig:stability}
\end{figure}

Finally, the stability of a high-current electron source is crucial for some applications. We consequently tested the current evolution for a few hours and confirmed that the cathodes under test were able to sustain the production of high-average currents with very low jitter; see Fig.~\ref{fig:stability}(a). A statistical analysis also indicates that typical relative rms fluctuation of $ \sigma_{\bar I} \equiv \mean{\bar{I}^2}^{1/2}/\mean{\bar{I}} \simeq 2$\% was achieved over six-hour periods and independently of the mean operating current $\mean{\bar{I}}$; see Fig.~\ref{fig:stability}(b,c).  \\
%

In summary we have demonstrated the operation of a CNT cathode in the pulsed regime and produced bunch trains with operating average current up to $\bar{I}=0.65$~A and duration of $\sigma_t\simeq 70$~ps implying a charge per bunch $Q \simeq \bar{I}/f_0 \simeq 0.50 $~nC corresponding to a single-bunch peak current $\hat{I} = Q/(\sqrt{2\pi}\sigma_t)\simeq 3$~A. The explored cold-cathode technology coupled with a superconducting resonator could lead to the development of high-average current quasi-continuous-wave electron sources. The main challenge toward such an endeavor remains the temporal control of the emission process as electrons field-emitted at unfavorable times are most likely to hit the resonator wall. Such collisions could result in secondary electron emissions and possible multipacting (as observed in some of our experiments) or could ultimately result in a superconducting quench of the cavity.  Therefore the development of gating schemes aimed at shortening the electron-bunch durations and preventing the back-propagation of electrons is crucial. A dual-frequency gun~\cite{lewellen} supporting a fundamental and harmonic frequencies could effectively gate the emission of the CNT cathode to the proper phase of the accelerating RF wave. Since the CNT cathodes have a distinct threshold voltage, unlike thermionic cathodes, the bunch duration could be made much shorter, eliminating the need for a bunching structure before injection into a subsequent accelerator. In such a scenario, it should be possible to reach $\sim10$-ps bunch durations. \\

We are grateful to D. P. Grote and J.-L. Vay for their help with {\sc warp}, to B. Chase, P. Prieto, E. Lopez and R. Kellett for  technical support and to E. Harms, S. Nagaitsev and V. Shiltsev for support. This work was funded via US Department of Energy (DOE) contract DE-SC0004459 with Radiabeam Technologies, LLC,  and executed under CRADA agreement FRA-2013-0006 between Fermilab  and Radiabeam Technology, LLC.  Fermilab is operated by the Fermi Research Alliance, LLC. for the DOE under contract DE-AC02-07CH11359.

\end{document}